# Quantum exciton solid with embedded electron-hole solids in double-layer WSe$_2$

Meizhen Huang,[1, #] Zefei Wu,[1, #,*] Chenxuan Lou,[1, #] S. T. Chui,[2, *] and Ning Wang[1, *]

[1]Department of Physics and State Key Laboratory of Optical Quantum Materials, The Hong Kong University of Science and Technology, Clear Water Bay, Hong Kong, China

[2]Bartol Research Institute and Department of Physics and Astronomy, University of Delaware, Newark, Delaware 19716, USA

[#]These authors contributed equally: Meizhen Huang, Zefei Wu, Chenxuan Lou.

*Corresponding author. Email: phwang@ust.hk; chui@udel.edu; zefeiw@gmail.com

**Abstract:** We studied double-layer WSe$_2$ stacked on opposite sides of thin layers of hexagonal Boron nitride with different densities of electrons and holes. For a fixed hole density, the Coulomb drag resistance is found to exhibit plateaus approximately equal to $-h/(4e^2)$ and $-h/(2e^2)$ as the electron density is changed. When the number of electrons is equal to the number of holes, an exciton solid forms whose transport of quantum edge defects gives rise to the drag resistance. When the electron and hole densities are different, the excess electrons form a solid embedded in the exciton solid. The Coulomb drag resistance of the exciton solid comes from the one-dimensional transport of the two lowest energy channels of quantum edge vacancy-interstitial pairs. This corresponds to the first plateau. With the embedded solid, one of these channels is blocked. This corresponds to the second plateau. Transport experiments in the Corbino geometry with no edges and extra heavier holes were carried out. The plateaus disappeared. Three peaks in the resistance at different hole densities were observed. We interpret that the three peaks correspond to the commensurate exciton and two classes of hole solids. We performed phonon calculations of these states and found that the stability of these exciton-based quantum solids shows good agreement with experiment. Our results establish classes of extreme quantum solid states, opening additional avenues for the study of strongly correlated quantum transport phenomena involving quantum defect states.

## I. Introduction

Most crystalline solids consist of arrays of ion cores with masses of the order of the proton mass. Quantum effects become important when the core mass becomes small. Thus, for solid $^4$He, the defects are expected to form waves with a large De Broglie wavelength at millikelvin temperatures. Supersolid behavior is proposed to come from these quantum defects [1-6]. For novel classes of solids of electrons and holes with masses that are a thousand times smaller, the effective masses are of the order of the electron mass [7]. Defects in these extreme quantum solids have the same De Broglie wavelength as electrons in ordinary solids and we expect new quantum physics to arise from the quantum defects in these systems. Solids of mixtures of $^3$He and $^4$He form under high pressure and many interesting behaviors are observed [8-11]. We found related extreme quantum solids of mixtures of electrons/holes and excitons when the electron and hole densities become unequal.

Indirect excitons, formed by spatially separated electron–hole pairs in coupled quantum wells or van der Waals heterostructures, provide a tunable platform for correlated phases due to their long lifetimes [12]. This tunability enables access to diverse arrays of interaction-driven phases with unique symmetry breaking, transport signatures, and collective excitations. A wide range of excitonic states have been theoretically predicted and experimentally observed over the past decades [13-17]. At high densities, strong screening leads to a metallic electron–hole plasma, while lower densities can induce a crossover to an exciton fluid or a BCS-like excitonic insulator, characterized by condensation and gap formation [17-26]. The difference between these two phases is the size of the pair relative to the distance between the pairs, analogous to the recently studied BEC-BCS transition in atomic Fermion systems [27]. Among these phases, the exciton solid represents a particularly intriguing but previously unconfirmed phase: a two-dimensional (2D) lattice of well-separated excitons [28,29]. A key property of this state is that quantum defects such as vacancies or interstitials, particularly near sample edges, can facilitate unconventional edge transport, offering a new conduction mechanism in correlated insulators.

Here we present clear experimental evidence for robust Coulomb drag plateaus in double-layer $WSe_2$ devices separated by a thin layer of hexagonal Boron nitride (hBN) with electrons on one layer and holes on the other layer controlled by dual-gating. By fixing the hole (electron) density in one layer and tuning the carrier densities of the opposite species independently in the other layer, we access regimes conducive to exciton solid formation and observe distinct drag plateaus at $-h/(4e^2)$ when the densities of the electrons and the holes are comparable and $-h/(2e^2)$ when there are excess particles. Crucially, in edge-free Corbino geometry, these plateaus are replaced by resistance peaks, confirming that transport is mediated by quantum defects propagating along the sample edges. Our results suggest the formation of the exciton solid when the densities of the two species in double-layer $WSe_2$ are comparable and then embedded electron (hole) solids when there are excess particles of one of the species. Supporting phonon spectrum calculations provided an estimate of the Lindemann's ratio that confirmed the stability of these phases against quantum melting. This research opens the door to classes of interesting physics involving extreme quantum solids.

## II. Coulomb drag measurements

Transition metal dichalcogenide (TMDC) heterostructures have emerged as an ideal platform for exploring strongly correlated excitonic phenomena. Their large effective masses, strong interlayer Coulomb coupling due to reduced dielectric screening, atomically sharp interfaces, and the capacity for independent control over carrier type and density in each layer [12,30], collectively enable the stabilization and experimental detection of long-predicted excitonic states. We investigated the Coulomb drag resistance in heterostructures composed of two few-layer $WSe_2$ sheets, spatially separated by a thin hBN insulating barrier (Fig. 1(a); for optical images of the device, see Supplementary Material Fig. S1 [31]). The device employs independent top and bottom gates ($V_{tg}$ and $V_{bg}$), allowing precise and independent control of carrier densities ($n$ for electron and $p$ for hole doping) in the two $WSe_2$ layers. Consistent with other Coulomb drag experiments[18,32], we chose the top layer as the drag layer because of its smaller contact resistance compared to that in the bottom layer. To probe interlayer Coulomb interactions, we use a standard drag measurement: an excitation current ($I_{drive}$) is applied to the drive layer, while the induced voltage ($V_{drag}$) is measured in the drag layer (Fig. 1(b)). Because the layers are electrically isolated by the hBN barrier, any voltage in the drag layer must arise from interlayer electron-electron or electron-hole momentum transfer. The Coulomb drag resistance is defined as $R_{drag} = V_{drag}/I_{drive}$, directly probing the many-body coupling between the layers. Unless otherwise specified, measurements were performed at a base temperature of $T = 1.5$ K. Throughout, $V_{bg}$ is held fixed at -70V, ensuring the bottom WSe$_2$ layer is robustly p-type (for a two-probe current versus gate curve for the bottom layer, see Supplementary Material Fig. S2 [31]), while sweeping $V_{tg}$ tunes the top layer from p-type (negative $V_{tg}$) to n-type (positive $V_{tg}$) (for a two-probe current versus gate curve for the top layer, see Supplementary Material Fig. S3 [31]), enabling both hole-hole (p-p) and electron-hole (n-p) regimes within a single device; fixing the bottom layer as p-type and tuning the top layer from p- to n-type via the top gate allows us to maintain low contact resistances.

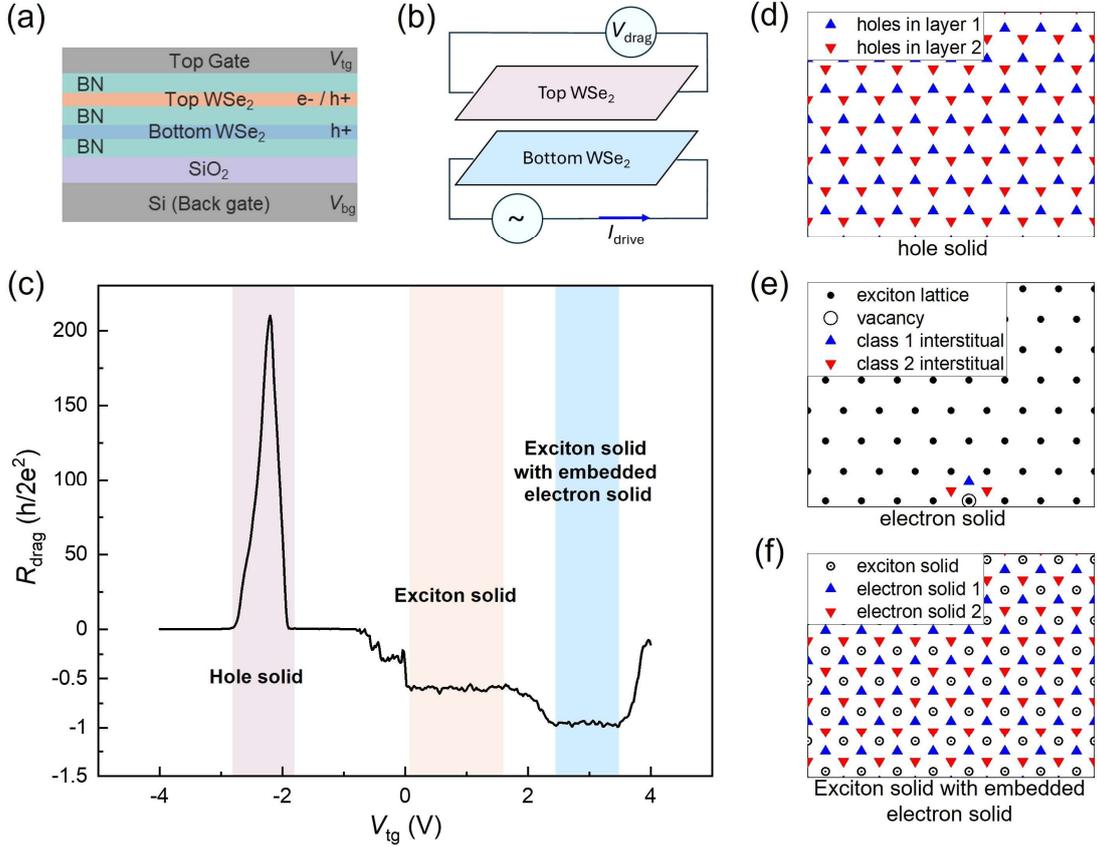

**FIG. 1.** Coulomb drag measurements in a Hall bar sample. **(a)** Schematic device structure. The SiO$_2$ thickness is 285 nm. (b) Schematic of Coulomb drag measurement in Hall bar sample. (c) Drag resistance vs $V_{tg}$ at $T = 1.5$ K. The system forms a hole solid with a sharp peak when $p_{top} = p_{bottom} \approx 3.2 \times 10^{12}$ cm$^{-2}$, forms an exciton solid with a plateau near $1/(4e^2/h)$ when $n_{top} = p_{bottom} \approx 2.6 \times 10^{12}$ cm$^{-2}$, and an exciton solid with one embedded electron solid with a plateau near $1/(2e^2/h)$ when $n_{top} \approx 4.8 \times 10^{12}$ cm$^{-2}$, $p_{bottom} \approx 2.4 \times 10^{12}$ cm$^{-2}$ so that $n_{top}: p_{bottom} = 2:1$. (d) Possible schematic of the hole solid, holes in two layers form hexagonal lattices. (e) Schematic of the exciton solid. The regular exciton lattice is indicated by the black dots. The vacancy is indicated

by the circle. Two types of low energy vacancy-interstitial pairs, indicted by the blue and red triangles, results in two quantum defect channels. (f) Schematic of the exciton solid with embedded electron solids. The number of channels in the transport becomes one when one of the two vacancy-interstitial quantum defect channels is blocked by the embedded electron solid.

Fig. 1(c) presents $R_{drag}$ versus $V_{tg}$. In the p-p regime, we observe a positive drag resistance with a pronounced peak when the hole densities are equal in both layers ($p_{top} = p_{bottom} \approx 3.2 \times 10^{12}$ cm$^{-2}$ at $V_{tg}$ = -2.2 V). This sharp peak is reminiscent of the electron solid we previously reported in double MoS$_2$ layers [33], and signals the formation of a strongly coupled correlated state—likely a "hole solid"—where long-range Coulomb interactions stabilize a crystalline arrangement of holes (see Fig. 1(d)). Sweeping $V_{tg}$ from negative to positive, major carriers of the top layer transition from holes to electrons, as indicated by the sign reversal in $R_{drag}$. In the electron-hole regime, $R_{drag}$ exhibits two distinct plateau regions near 1/(4e$^2$/h) and 1/(2e$^2$/h), respectively. These robust plateaus persist over a range of gate voltages and carrier densities, indicating the emergence of novel collective ground states. The plateau values are robust and independent of the bottom-gate voltage (for drag resistance versus $V_{tg}$ curves at different $V_{bg}$, see Supplementary Material Fig. S4 [31]), confirming that they are intrinsic physical phenomena and not measurement artifacts. In the following sections, we establish that these plateaus originate from the formation of a quantum exciton solid and a composite phase with an embedded electron solid.

### III. Exciton solid with drag resistance near 1/(4e$^2$/h)

We first focus on the resistance plateau near 1/(4e$^2$/h) at $V_{tg} \approx$ 1V, where electron and hole densities in the two layers are matched ($n_{top} = p_{bottom} \approx 2.6 \times 10^{12}$ cm$^{-2}$ at $V_{tg} \approx$ 1.3 V) and are confined to opposite sides of a hBN spacer ($d \approx$ 5 nm). Our prior theoretical calculations [28] indicate that for typical TMDC parameters, the exciton transverse size ($\xi \approx$ 23 Å) is much smaller than the interexciton spacing ($a \approx$ 100 Å at densities ~10$^{12}$ cm$^{-2}$), confirming well-defined, nonoverlapping excitons. However, dipole-dipole repulsion alone cannot stabilize a quantum solid; the Lindemann ratio is ~20%, implying melting via dislocation unbinding. This is consistent with previous theoretical estimates of the exciton Mott density, which considered unpinned exciton fluids and did not include the periodic potential of the hBN substrate [12]. Once the hBN charge-transfer potential is incorporated, our self-consistent phonon calculations [28] and fixed-node diffusion Monte Carlo simulations [26] confirm the dynamical stability of the exciton solid. This stability originates from the hBN substrate, where periodic charge transfer between boron and nitrogen atoms ($Q \approx 0.47e$) [34] imposes a static periodic potential that pins the exciton lattice. This pinning suppresses quantum fluctuations and reduces the Lindemann ratio to ~7%, thereby enabling stability at low temperatures. As excitons are electron–hole bound states, the current and voltage responses are linked: the drag resistance has the same magnitude as the ordinary resistance, but with the opposite sign due to the charges of the constituent particles.

The exciton solid is a bulk insulator with excitons localized on a triangular lattice, leaving no itinerant charge carriers, distinct from an excitonic insulator [15,35], charge density wave [36] (where Fermi surface nesting is required), or electron-hole plasma. Nevertheless, its defects (vacancies, interstitials or dislocations) have effective masses comparable to electrons/holes [7], much less than that of atoms in ordinary solids. yielding De Broglie wavelengths three orders larger than in atomic solids, rendering them intrinsically quantum. The quantum motion of defects is central to the proposed supersolid-like behavior in solid $^4$He. These defects readily form at the sample edges, and their one-dimensional propagation along the perimeter enables edge conduction. According to Landauer [37], the conductance of a one-dimensional quantum channel is $G = MPG_0$, with $G_0 = 2e^2/h$ the conductance quantum, $P$ the transmission probability, and $M$ the number of transport channels. In one dimension, the distinction between impenetrable bosons and fermions disappears [38]. For non-penetrating bosons, the wavefunction signs in regions between particles are uncorrelated, resulting in effective fermionic statistics. This mapping is independent of the spin of the constituent particles; the exclusion principle is enforced by strong interparticle repulsion rather than spin degrees of freedom. Consequently, the spin does not contribute additional transport channels. Given that the transmission is not perfect ($P < 1$), the Coulomb drag resistance satisfies $-R_M > 0.5/(Mh/e^2)$. Experimentally, the first resistance plateau (Fig. 1(c)) near $1/(4e^2/h)$ corresponds to $M = 2$ indicating two defect channels. This is also illustrated in Fig. 1(e), where two types of low-energy vacancy-interstitial pairs each support a transport channel. Thus, the exciton solid supports two quantum channels for defect-mediated edge transport.

**IV. Exciton solid with embedded electron solid with drag resistance near $1/(2e^2/h)$**

Beyond the first plateau, increasing $V_{tg}$ further reveals a second plateau in the drag resistance near near $1/(2e^2/h)$. This plateau appears at $V_{tg} \approx 2.4$ V, corresponding to $n_{top} \approx 4.8 \times 10^{12}$ cm$^{-2}$ and $p_{bottom} \approx 2.4 \times 10^{12}$ cm$^{-2}$ with an electron-to-hole density ratio of approximately 2:1. We attribute this to the formation of an electron solid embedded within the existing exciton solid matrix (Fig. 1(f)). This occurs as additional electrons accumulate and self-organize into a periodic structure coexisting with the exciton lattice. The embedded electron solid can adopt two symmetry-equivalent configurations ("solid 1" and "solid 2") relative to the exciton lattice, both energetically degenerate due to the underlying lattice symmetry. In our drag measurements, a plateau slightly above $1/(2e^2/h)$ is observed, indicating transport through only one conducting channel. This reduction is explained by the blocking of one of the two quantum defect channels (vacancy-interstitial channels) by the presence of the embedded electron solid, effectively suppressing its contribution to the total conductance. The resulting single-channel conduction is consistent with both the quantized resistance value and our many-body interpretation.

**V. Experimental Results: Geometry Dependence**

To provide support that the plateau comes from edge transport, we performed measurements using Corbino-geometry devices (Fig. 2(a), for a detailed structure of the Corbino device, see Supplemental Material Fig. S5 [31]), which eliminate physical edges. With the top layer fixed as p-type, we tuned the bottom layer continuously from p- to n-type. Strikingly, in the p-n region, the drag resistance exhibited three distinct peaks instead of plateaus (Fig. 2(b)). The absence of the plateau in this geometry confirms that its formation requires edge-state transport.

While mechanisms like density-dependent screening, percolation, or electrostatic inhomogeneity can influence transport, they are unlikely to produce these specific, discrete peaks. For example, an increase of resistance can come from weak localization (quantum percolation) in two dimensional systems [39]. However, as the density is changed, there is no abrupt increase in the resistance from this mechanism. Similarly, the formation of a collection of exciton from an electron hole plasma can be thought of as a change of screening when the number of electrons is commensurate with the number of holes [40], and could in principle modify the resistance. However, we have recently performed fixed node quantum diffusion Monte Carlo simulation for the geometry of our experimental structure and found that excitons are already favorable for the range of densities in the current paper [26]. Thus, no abrupt change in resistance is expected from this mechanism either. Meanwhile, there can be electrostatic inhomogeneity in the sample but the low resistance at low densities shows that there are continuous connected paths along which charges can move. The lattice spacing of the lattice is of the order of 100 Angstrom. We do not expect an abrupt change in the available conducting paths as the particle density is changed.

We interpret these three peaks as signatures of composite many-body states arising from the interplay between excitons and additional holes in the top layer: specifically, the three peaks from left to right correspond to (1) an exciton solid embedded with two hole solids ($p_{top} \approx 6.50 \times 10^{12}$ cm$^{-2}$, $n_{bottom} \approx 2.13 \times 10^{12}$ cm$^{-2}$), (2) an exciton solid with one embedded hole solid ($p_{top} \approx 4.51 \times 10^{12}$ cm$^{-2}$, $n_{bottom} \approx 2.26 \times 10^{12}$ cm$^{-2}$), and (3) a bare exciton solid ($p_{top} = n_{bottom} \approx 2.45 \times 10^{12}$ cm$^{-2}$). This sequence reflects the gradual reduction of hole solids as the electron density in the bottom layer increases. The absence of quantized plateaus in the Corbino geometry, in contrast to the Hall bar geometry, reveals the crucial role of edge states and their associated defects in stabilizing the drag resistance plateaus, supporting our model of edge-mediated conduction in exciton quantum solids.

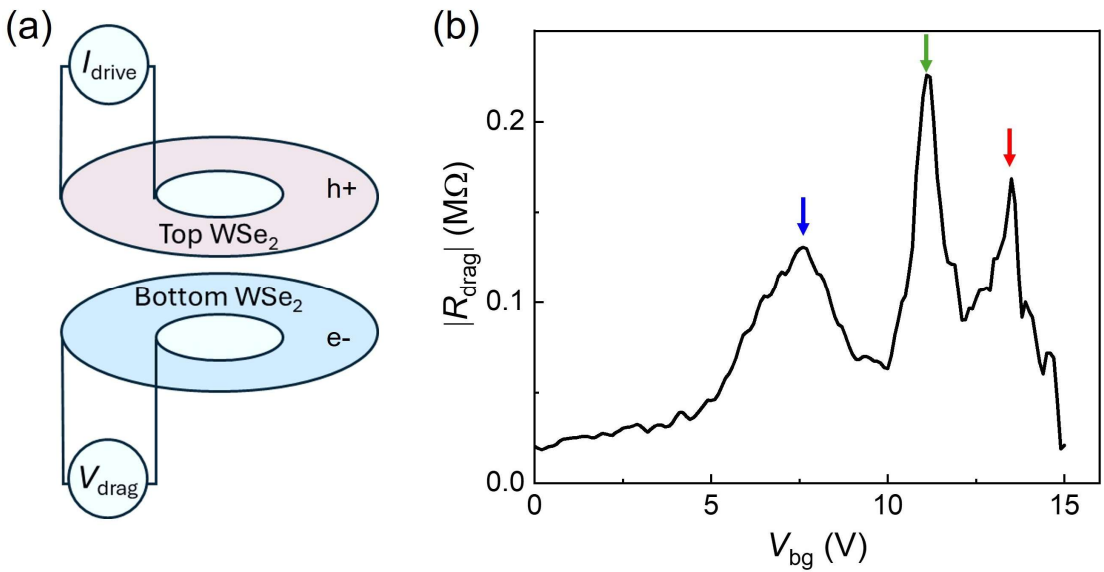

**FIG. 2.** Coulomb drag measurements in a Corbino sample. (a)Schematic of Coulomb drag

measurement in edge-free Corbino device. (b) Absolute value of the drag resistance vs $V_{bg}$. Instead of resistance plateaus, the drag resistance shows three peaks corresponds to an exciton solid embedded with two hole solids (blue arrow), an exciton solid with one embedded hole solid (green arrow), and a bare exciton solid (red arrow).

**VI. Phonon Analysis and Stability**

This work reports the transport signature of an exciton solid phase with embedded electron or hole solids. We systematically probed the stability and physical properties of these states. In our experiments, extra charges are introduced either as electrons (in the regular Hall bar geometry, with hBN thickness 5 nm) or as holes (in the Corbino geometry, with hBN thickness 3 nm). Building on prior calculations for a two-layer electron system [41] and electron-hole systems [28], we extend the phonon mode analysis to the exciton solid phase with embedded charge solids, incorporating Coulomb interactions between charges and the screening arising from polarization charges at the dielectric interfaces [26]. In addition, the stabilizing potential from the hBN substrate, previously discussed for the exciton solid [28], is also applied to the embedded solid states. As noted earlier, because the period of the substrate potential is much smaller than the average interparticle spacing, the particles are positioned at the potential minima, due to a slight distortion of the static particle position.

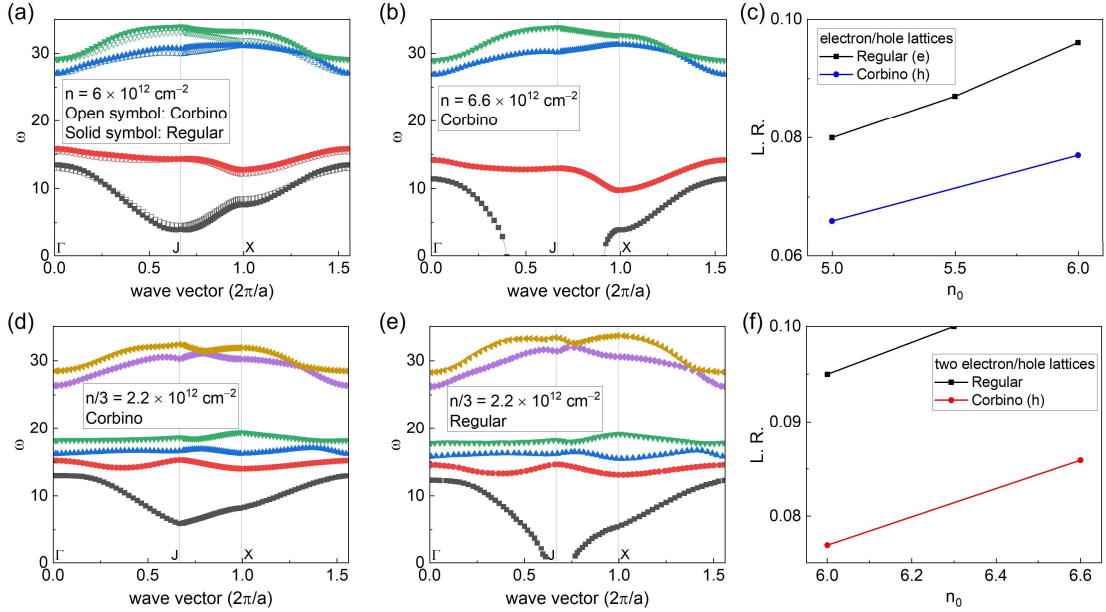

**FIG. 3.** Phonon analysis for exciton solid and embedded solid. a-c, phonon spectrum ((a) and (b) for different densities) and Lindemann's ratio at $n=n_0$ (c) for an exciton solid embedded with one electron/hole solid. For the Corbino (Regular) geometry $m_{eff}$ is the effective mass of the hole (electron). n denotes the carrier density in the layer of $WSe_2$ with a higher density. The value n/2 (n/3) corresponds to the density of the exciton solid and the charged solid when there is one (two) charged solid embedded in exciton solid. d-f, phonon spectrum ((d) and (e)) and Lindemann's ratio (f) for an exciton solid embedded with two electron/hole solids.

The resulting phonon spectrum for the exciton solid with one embedded solid (Fig. 3(a)) reveals two distinct groups of frequencies: the higher two branches correspond to the exciton solid, while the lower two are associated with the embedded electron or hole solid. In this analysis, we have neglected any out-of-plane motion of the particles, focusing only on in-plane dynamics. Notably, the stability of the embedded state is robust under experimental conditions, and variation in hBN thickness has little effects on phonon frequencies since the exciton solid is stabilized by the electrostatic potential from boron and the nitrogen ions at the interfaces with the hBN from the spacer and the encapsulation [28]. The unit of the angular frequency is inversely proportional to the inverse square root of the effective mass $m_{eff}$. Since the hole effective mass is about twice that of the electron, the phonon frequency for the Corbino geometry (with holes) is higher.

At higher density, however, the embedded state becomes locally unstable, as shown in Fig. 3(b). We evaluated its stability using the generalized Lindemann ratio. The relative mean square lattice vibration is $\langle r^2 \rangle = \sum_q \hbar(2n_q + 1)f_q/m\omega_q$ where $n_q$ is the number of phonons, $f_q = 1 - \cos(\boldsymbol{q} \cdot \boldsymbol{r}_0)$ where $\boldsymbol{r}_0$ is the position of a nearest neighbour. The Bohr radius is $a_B = \hbar/(me^2)$. From the phonon frequency we first calculate the "bare" mean-square lattice fluctuation

$$r_0^2 = \sum_q f_q / \omega_q.$$

The Lindemann ratio is given by

$$\langle \delta r^2 \rangle^{1/2} / a = (a_B / a)^{1/4} r_0.$$

This relative Lindemann's ratio is shown in Fig. 3(c) as a function of the density. Empirically, quantum melting occurs when this ratio is about 10 % [42,43]. Our result shows that this state has not melted over a region of experimental interest, while the exciton solid with an embedded electron solid is less stable due to the smaller electron mass and larger quantum fluctuations.

For the case of two embedded electron (or hole) solids at density of $n = 6.6 \times 10^{12} cm^2$, phonon spectra for holes (Corbino geometry, Fig. 3(d)) and electrons (regular geometry, Fig. 3(e)) indicate that the state is locally stable only for holes. The corresponding Lindemann ratio (Fig. 3(f)) approaches 10% for two embedded electron solids, suggesting that this state is not observed experimentally in the regular geometry because it has already melted.

**VII. Quantum melting at high temperatures**

To further investigate the thermal stability of these solid phases, we measured the temperature dependence of the drag resistance (see Fig. 4). In the Hall bar geometry, for both the exciton solid and the exciton solid with an embedded electron solid, the drag resistance remains nearly unchanged and the plateau constants up to 50 K (Fig. 4(a) and 4(b)), indicating that both phases are robust against thermal fluctuations at low temperature. Above 50 K, the resistance deviates, marking the onset of quantum melting.

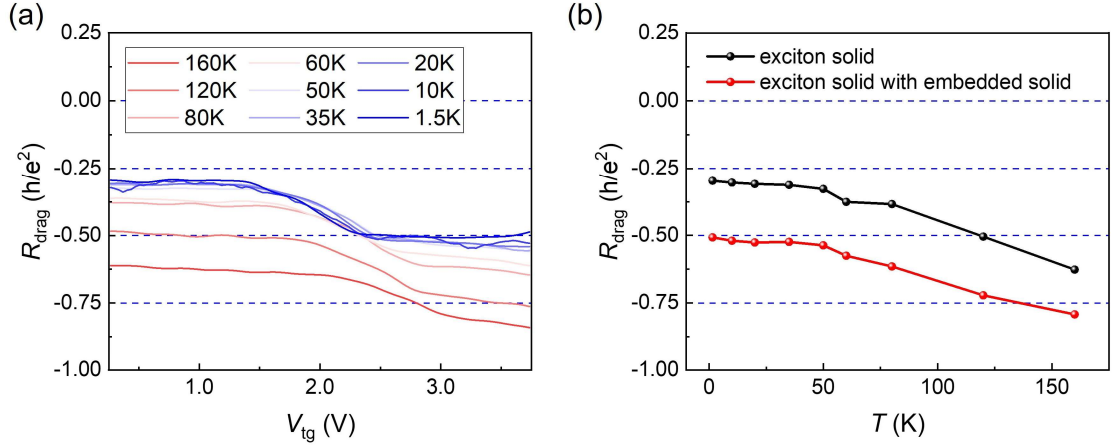

**FIG. 4.** Quantum melting of the solid states at high temperature. (a) Temperature dependence of drag resistance for an exciton solid and an exciton solid with an embedded electron solid. (b) Drag resistance vs temperature for $V_{tg} = 1$ V (black) and $V_{tg} = 3$ V (red).

## VIII. Conclusion

In summary, we report robust Coulomb drag plateaus in double-layer $WSe_2$, evidencing stabilized quantum solid phases of interlayer excitons and their composites. By tuning carrier type and density in dual-gated, atomically thin devices, we observe both a pure exciton solid and a novel exciton solid embedded with a commensurate electron (or hole) solid. The robustness of these phases is supported by multiple independent lines of evidence. First, the key signature of the exciton solid—quantized drag plateaus in Hall bar geometry—is replaced by three well-defined resistance peaks in an independently fabricated Corbino device (Fig. 2), consistent with the edge-transport model. Second, the plateau values remain unchanged under different fixed bottom-gate voltages (see a plot at Supplementary Material Fig. S4 [31]), ruling out accidental pinning by specific electrostatic conditions. Third, our phonon calculations (Fig. 3), which incorporate substrate-induced commensurate pinning, demonstrate that the proposed solid phases are energetically stable over a finite density range, with Lindemann ratios well below the quantum melting threshold. Taken together, this complementary experimental and theoretical evidence confirms that the observed phenomena are intrinsic to the underlying physics rather than device-dependent particulars.

The interplay between long-range dipolar interactions, substrate commensurability, and many-body quantum fluctuations emerges as a key ingredient for realizing and stabilizing such intricate states of matter. Notably, embedding charge solids further enables the design of composite quantum matter with tunable properties. Our findings resolve long-standing questions on bosonic ground states and establish TMDC heterostructures as a versatile platform for exploring correlated and topological phenomena in two dimensions. This integration of material control and theory opens pathways to new forms of quantum matter with fundamental and technological relevance.

**Acknowledgements**

Grant support from the National Key R&D Program of China (2020YFA0309600) and the Research Grants Council (RGC) of Hong Kong (Project Nos. 16302621, AoE/P701/20, C6053-23G-D) are acknowledged. This research work is partially supported by the SKL of Optical Quantum Materials, HKU. Project No.: SKLOQM26SC01. We also acknowledge Mr. Chun Kit Lai, Mr. Gordon C.T. Suen, and Dr. Yuan Cai for their valuable technical support during device fabrication at the MCPF and WMINST of HKUST.